\documentclass[pra,twocolumn,floatfix,showpacs,preprintnumbers,groupeaddress,amsmath,amssymb]{revtex4-1}

\usepackage[utf8]{inputenc}
\usepackage{mathptmx}
\usepackage{psfrag}
\usepackage{graphicx}						
\usepackage{braket}
\usepackage{mathtools}

\usepackage{color}
\definecolor{dred}{rgb}{.6,.0,0.}
\definecolor{dblue}{rgb}{.0,.0,0.6}

\usepackage{tikz}
\usetikzlibrary{shapes.geometric}

\colorlet{mfarbe}{magenta}

\begin{document}

\title{Quantum properties of a strongly driven Josephson junction}

\author{Jennifer Gosner}
\author{Bj\"orn Kubala}
\author{Joachim Ankerhold}
\affiliation{ Institute for Complex Quantum Systems and IQST, University of Ulm, 89069 Ulm, Germany}

\date{December 6, 2018}

\begin{abstract}
A Josephson junction embedded  in a dissipative circuit can be externally driven to induce nonlinear dynamics of its phase. Classically,  under sufficiently strong driving and weak damping, dynamic \emph{multi-stability} emerges associated with dynamical bifurcations so that the often used modeling as a Duffing oscillator, which can exhibit \emph{bi-stability} at the most, is insufficient. The present work analyzes in this regime corresponding quantum properties by mapping the problem onto a highly-nonlinear quasi-energy operator in a rotating frame. This allows us to identify in detail parameter regions where simplifications such as the Duffing approximation are valid, to explore classical-quantum correspondences, and to study how quantum fluctuations impact the effective junction parameters as well as the dynamics around higher amplitude classical fixed points. 
\end{abstract}

\maketitle

\section{Introduction} 
Nonlinear quantum oscillators driven by external time-periodic forces have received substantial attention in the last decade,  particularly in nanomechanical systems and mesoscopic devices, see \cite{Dykmanbook2012,PootvanderZant2012} and references therein. These activities strive on the one hand to obtain an understanding of fundamental properties of quantum-classical transitions \cite{MartinisDevoretClarke1987,GROSSMAN1993295,GuoPeanoMarthalerDykman2013} and on the other hand to make use of generic nonlinear phenomena such as bifurcations \cite{SiddiqiVijayPWMRFDevoret2004, SiddiqiVijayVionEsteve2005,ManucharyanBMVijaySiddiqiDevoret2007, VijayDevoretSiddiqi2009,PhysRevLett.106.094102,
PhysRevB.83.224506} in actual implementations, e.g.\  for enhanced sensing schemes. In many of these situations theoretical descriptions rely on the paradigmatic case of a Duffing oscillator, a driven damped oscillator with quadratic plus quartic potential surface, as long as anharmonicities are sufficiently weak to treat them perturbatively \cite{DykmanKrivoglaz1980,PeanoThorwart20062,SerbanWilhelm2007,GuoZhengLi2010, GuoZhengLiZhengLiYan2011,andreguopeanoschoen2012}. In this way, quantum-classical transitions have been analyzed within Wigner function approaches \cite{Kheruntyan1999,KatzRetzkerStraubLifshitz2007, GuoZhengLi2010,GuoZhengLiZhengLiYan2011,GevorgyanKryuchkyan2013} and the seminal phenomenon of quantum activation has been revealed \cite{MarthalerDykmanswitchin2006, DykmanMarthalerPeano2011,PhysRevLett.110.047001}. Experimentally, nonlinearities appear naturally in {\em the} key element of superconducting circuits, namely, Josephson junctions. Such circuits are used in form of Cooper pair boxes to implement artificial atoms in circuit quantum electrodynamics \cite{NakamuraPashkinTsai1999, VionAassimeCottetJoyezPothierUrbinaEsteveDevoret2002,
WallraffSchusterBlaisSchoelkopf2004}, but are also exploited in Josephson bifurcation amplifiers as quantum limited detectors \cite{SiddiqiVijayPWMRFDevoret2004, SiddiqiVijayVionEsteve2005, VijayDevoretSiddiqi2009}. The nonlinear quantum properties of Josephson junctions are essential in the new field of Josephson photonics, where non-classical radiation is created by a dc-voltage biased junction placed in series with a single or multiple resonators \cite{HofheinzPortierBaudouinJoyezVionBertetRocheEsteve2011,PhysRevB.90.020506, Cassidy939,WestigKubalaPortier2017,1810.06217}. In this latter application, the light-matter coupling constant can reach easily relatively large values compared to conventional implementations, so that nonlinear effects can become extremely pronounced \cite{1810.06217}. The full nonlinearity gives rise to a wealth of quantum dynamical features far from equilibrium which may offer new platforms to access phenomena such as critical slowing down or Kibble-Zurek-type of scenarios \cite{KIBBLE1980183,ZUREK1996177, PhysRevLett.95.105701}. It is also important for current research which extends concepts from the pair-creation resonance \cite{DykmanMaloneySilverstein1998,PhysRevB.83.224506,DykmanMarthalerPeano2011} to higher-order photon resonances \cite{PhysRevLett.111.205303,1367-2630-18-2-023006,1742-6596-681-1-012018,ZhangGosnerDykman2017}  with the ultimate goal to create multi-cat states and ensembles of entangled photons \cite{Vlastakis607,MirrahimiJiangDevoret2014,Goto2016, SvenssonShumeikoDelsing2017,180209259}. 

Very recently, in an extension of the dynamical regime, where the Josephson bifurcation amplifier operates, the strong driving domain of a Josephson device, realized in form of a Superconducting Interference Device (SQUID),  with the phase as the relevant classical degree of freedom has been studied \cite{JungUstinov2014}. As expected, higher order anharmonicities give rise to a multitude of steady state orbits with  a complex bifurcation pattern. The motivation of this present work is to explore aspects of these phenomena in the quantum realm by considering a Josephson junction with its phase being quantum and subject to an external  AC-current drive such that the Duffing approximation fails. 
By mapping the Hamiltonian from the laboratory frame to a frame rotating with the frequency of the external source, we arrive at a quasi-energy operator that is analyzed both in absence and in presence of dissipation. This allows identifying regimes, where simplifications such as the Duffing oscillator are valid. A comparison between classical fixed points and quantum steady state distributions in Fock space reveals possible classical-quantum correspondences. Note that the set-up we consider here is very different from the voltage driven case, where the simultaneous presence of a dc- and an ac-voltage gives rise to the well-known Shapiro resonances \cite{Shapiro1963}. We also consider the role of local quantum fluctuations and squeezing close and away from classical instabilities which  may experimentally appear as voltage fluctuations. This may motivate further experiments to explore the wealth of the nonlinear quantum dynamics in Josephson junctions.

The paper is organized as follows: In Sec.~\ref{hamilton} the model and its mapping to the rotating frame together with simplifications is introduced. The corresponding classical system is briefly addressed in Sec.~\ref{Classsys}, before in Sec.~\ref{barehamilton} we analyze the energy spectra of the undamped system. The remainder, Sec.~\ref{steadystate},  is devoted to the steady state behavior based on a quantum master equation. Conclusions are drawn in Sec.~\ref{conclusion}.

 \section{Time-periodic driving and the Josephson Hamiltonian}\label{hamilton}
\begin{figure}
 \includegraphics[width=1\linewidth]{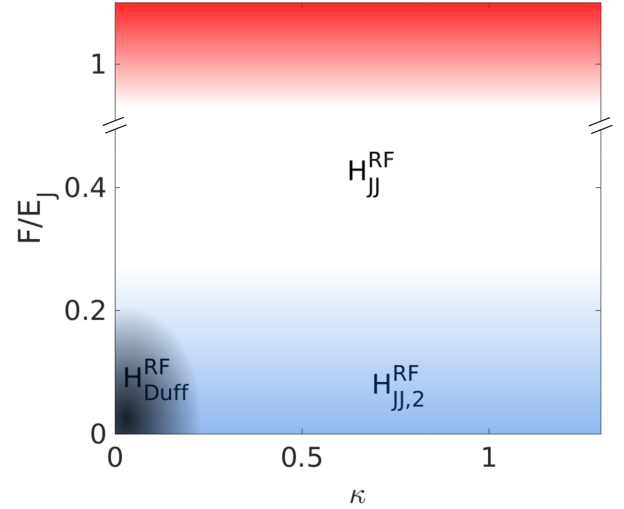}
 \caption{Parameter regimes of the Josephson Hamiltonian and regions of validity of different approximations in a rotating frame. The rotating wave approximation fails for strong driving, $F/E_J \gtrsim 1$ (red). 
For weak driving, $F/E_J \ll \tilde{\gamma}$, the Bessel function in 
(\ref{hasfje}) can be expanded to the lowest order nonlinear term, so that  $ {H}_\text{JJ}^\text{RF} \approx {H}_\text{JJ,2}^\text{RF}$ (blue, plotted for $\tilde{\gamma}=0.2$). Conventional descriptions employ the Duffing Hamiltonian (\ref{duffHammm}), which further requires $\kappa \ll 1$. 
With increasing $\kappa$ the system moves from a classical towards a quantum regime. }
 \label{classdegg}
\end{figure}

A Josephson junction with phase difference $\varphi$ and conjugate charge operator  $q$ subject to a time-periodic driving  is described by
\begin{equation}
H_{\text{JJ}}= \frac{q^2}{2 C} - E_J \cos(\varphi) - F \cos(\omega_d t) \varphi\, 
\label{werwtta}
\end{equation}
with $[{q},{\varphi}]= - i 2e\hbar $ so that formally $q/2e$ can be seen as the momentum operator conjugate to the position $\varphi$. The driving amplitude is $F=\phi_0 I_d$, where $\phi_0$ is the reduced flux quantum and $I_d$ denotes the maximum of the bias current.
\newline
Now, in a first step we transform the above Hamiltonian to a frame rotating with the frequency of the external drive, i.e., ${H}_{}^{\text{RF}}={U}{H}{U}^{\dagger} + i \hbar {U}^{\dagger}\partial_t  {U}$ with the unitary operator ${U}=e^{i \omega_{\text{d}} t {a}^{\dagger}{a}} $ (similar to \cite{MarthalerDykmanswitchin2006}). Here $a, a^\dagger$ denote canonical annihilation and creation operators, respectively, according to $\varphi= \kappa (a+a^\dagger)$ with $[a,a^\dagger]=1$ and ground state width in the Josephson potential $\kappa^2=\sqrt{E_C/2 E_J}$  with charging energy $E_C=2 e^2/C$. In terms of a mechanical analog, this corresponds to the ground state width of a harmonic system with frequency $\omega=\sqrt{2 E_J E_C}/\hbar$ and mass $M= \hbar^2/2 E_C$, i.e. $\kappa=\sqrt{\hbar/ 2M\omega}$. Dropping all fast oscillating terms, a straightforward calculation then leads to the Hamiltonian (\ref{werwtta}) in the rotating wave approximation, i.e., 
\begin{equation}
{H}_{\text{JJ}}^{\text{RF}}=   \hbar \delta\!\omega\,  \hat{n}- \frac{F}{2} \kappa({a}+{a}^{\dagger})  
 -  {E}_J^* \left[ :\hspace{-0.05cm}J_0\left( 2 \kappa \sqrt{ \hat{n}} \right)\hspace{-0.05cm}: + \kappa^2 \hat{n} \right] \,,\!
\label{hasfje}
\end{equation}
where  $\hat{n}=a^\dagger a$ and  $J_0$ is the zeroth-order Bessel function with the colons indicating normal ordering. Further, one has 
 $\delta\!\omega={\omega}^*-\omega_d$ with the renormalized plasma frequency ${\omega}^*= \omega\,  (1+ {\rm e}^{-\kappa^2/2})/2 $ and the renormalized Josephson energy
$E_{\text{J}}^*= E_{\text{J}} {\rm e}^{-\kappa^2/2}$ (cf. \cite{Joyez2013,1810.06217}). 
The rotating wave approximation limits the validity of this Hamiltonian to a domain of small detuning, $|\delta\!\omega|\ll \omega_d$, and also sets an upper limit to the driving amplitude, see Fig.~\ref{classdegg}. The latter stems from requiring that the energy level spacing of the autonomous system sufficiently exceeds the energy scale for driving, i.e.,\ $\kappa F< \hbar\omega$ which is equivalent to $F/E_J< 1$.

It is instructive to consider the situation, where $\kappa \sqrt{\langle \hat{n}\rangle}\ll 1$ so that the Bessel function in (\ref{hasfje}) can be expanded. Expressed in terms of the typical steady state value for  harmonic systems
at resonance $\langle \hat{n}\rangle\sim (\kappa F/\hbar\gamma)^2$ with damping rate $\gamma$, this implies $(F/E_J) \ll \tilde{\gamma}$, where $\tilde{\gamma}=\gamma/\omega $ is the inverse of the Q-factor of the junction at the bare plasma frequency. This way, one finds
\begin{equation}
{H}_{\text{JJ, 2}}^{\text{RF}}=\hbar \delta\!\omega\,  \hat{n} -E_J^*  \frac{\kappa^4}{4} ( \hat{n}^2 -  \hat{n})
- \frac{F}{2} \kappa ({a}+{a}^{\dagger})\, ,
\label{waermbncjtwo}
\end{equation}
where we kept the lowest order nonlinear term in the occupation operator. 

If now  $\kappa\ll 1$ meaning that $E_J$ by far exceeds $E_C$, the above result can be further simplified to
\begin{equation}
{H}_{\text{Duff}}^{\text{RF}} =  \hbar \delta\!\omega_0\, \hat{n}  - E_{\text{J}}  \frac{\kappa^4}{4} \left( \hat{n}^2 +  \hat{n} \right) - \frac{F}{2} \kappa ({a}+{a}^{\dagger})\, ,
\label{duffHammm}
\end{equation}
where $\delta\!\omega_0=\omega-\omega_d$. This expression coincides with the rotating-wave approximated form of the Duffing oscillator Hamiltonian that has been used in various contexts recently \cite{andreguopeanoschoen2012, SerbanWilhelm2007,GuoZhengLi2010, GuoZhengLiZhengLiYan2011}. We recall that the latter describes a periodically driven  nonlinear oscillator with purely quartic anharmonicity that typically is assumed to carry a parameter (length scale) that can be tuned independently from the harmonic part of the potential. This, however, is not the case for the Josephson junction, where $E_J$ serves as the single parameter which controls both the plasma frequency at low amplitudes as well as the anharmonicity length scale.

It is thus the domain outside the range of validity of Eqs.~(\ref{waermbncjtwo}), (\ref{duffHammm}) where a deep quantum regime can be found, see Fig.~\ref{classdegg}. There, ground state fluctuations of the phase (seen as a collective degree of freedom of the circuit) influence the effective junction parameters such as plasma frequency and Josephson coupling. Formally,  in the  limit of  large $\kappa > 1$ (but still $\kappa^2\langle n\rangle\ll 1$ which implies $F/E_J\ll 1$) the expression (\ref{waermbncjtwo}) can be written as a Hamiltonian of a harmonic system, i.e., ${H}_{\text{JJ, 2}}^{\text{RF}}\to \hbar (\omega/2-\omega_d)\, \hat{n} -F \varphi$, with resonance frequency  $\omega/2$.

To analyze properties of the driven Josephson system in more detail in the following, it is convenient to introduce dimensionless quantities. For this purpose, we measure all energies in units of $M\omega\delta\!\omega$ and further introduce
\begin{equation}
f=\frac{F}{M\omega\delta\!\omega}\ , \ \epsilon=\frac{E_J^*}{M\omega\delta\!\omega}\, 
 \end{equation}
 as dimensionless parameters. In these units and expressed in dimensionless position and momentum operators $\hat{Q}$ and $\hat{P}$, respectively, according to $a=(\hat{Q}+i \hat{P})/(2\kappa)$ and with $[\hat{Q}, \hat{P}]=2 i\kappa^2$, one obtains  $g\equiv H_{\text{JJ}}^{\text{RF}}/(M\omega\delta\!\omega)$ as
\begin{equation}
\begin{split}
g(\hat{Q},\hat{P})= &\phantom{-}\frac{1}{2} \left( \hat{Q}^2 + \hat{P}^2 \right)- \frac{f}{2} \hat{Q}  \\
&- \epsilon\, \left[\, : J_{0}\left({\sqrt{\hat{Q}^2+\hat{P}^2}} \right):\,+\,\frac{1}{4} (\hat{Q}^2+\hat{P}^2)\right] \, .
\end{split}
  \label{clhamrotpot}
\end{equation}
Likewise, the quasi-energy operator for the Duffing system corresponding to \eqref{duffHammm} and also scaled with $M \omega \delta \omega$ reads
\begin{equation}
\begin{split}
g_{\rm Duff}(\hat{Q},\hat{P})= &\left(\frac{\delta\!\omega_0}{2 \delta\!\omega}- \frac{\epsilon_0 \kappa^2}{16}\right) \left( \hat{Q}^2 + \hat{P}^2 \right)- \frac{f}{2} \hat{Q} \\ 
&- \frac{\epsilon_0}{64}\,   \left(\hat{Q}^2+\hat{P}^2\right)^2  \, ,
\end{split}
 \label{quasiduffing}
\end{equation}
where $E_J^*$ is replaced by $E_J$ in $\epsilon_0$.


\section{Classical system}  
\label{Classsys}
\begin{figure} 
\includegraphics[width=1\linewidth]{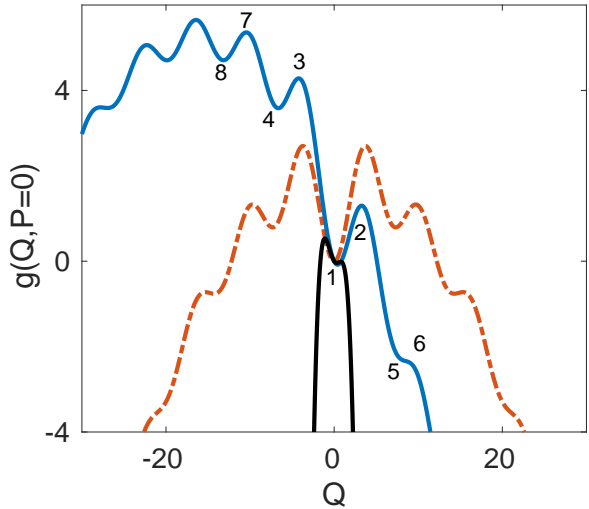} 
\caption{
Classical limit of the Hamiltonian $g(Q,P=0)$ for $\epsilon=17.3$, $f=0.56$ (black solid line), and for $\epsilon=2.05$ with ($f=0.8$, blue solid line) and without ($f=0$, red dashed line) driving. 
Stable orbits (for small damping) are found for all minima of quasi-energies at $Q \geq 0$ and for all maxima at $Q<0$ (labeled by odd numbers in increasing order of amplitude, cf.~Fig.~\ref{regimestatdyntwo}). 
All maxima (minima) of $g(Q,P)$  for $Q>0$ ($Q<0$) correspond to unstable solutions (even labels). The case $\epsilon \gg 1 $ corresponds locally to a Duffing oscillator, while for $\epsilon\gtrsim 2$ the full impact of the Bessel function is seen. In this regime multiple stable solutions can be found even without driving. Driving breaks the $Q-P$ symmetry and allows to experimentally access higher dynamical stable solutions.}
\label{regimestatdyn}
\end{figure}

In order to explore quantum-classical correspondences later, we here present a brief analysis of the classical steady state dynamics of the Hamiltonian (\ref{clhamrotpot}).  A full classical limit  consists  in replacing operators by classical phase-space variables in the rotating frame and setting $\kappa=0$ so that parameters reduce to bare values, i.e.\ $\delta\!\omega=\delta\!\omega_0$, $E_J^*=E_J$.
Qualitatively, one identifies a limit of small $Q^2+P^2$, where an expansion of the Bessel function  in the last line of 
(\ref{clhamrotpot}) leads to a dominating nonlinearity of the form $\frac{1}{2} (Q^2+P^2)^2$, so that the classical Duffing result is regained, i.e., putting $\delta\!\omega_0/\delta\!\omega=1$ and $\kappa=0$ in (\ref{quasiduffing}). In the opposite limit of large $Q^2+P^2$, the Bessel function tends to zero so that the quasi-energy turns into a harmonic function with linear tilt, where asymptotically $g(Q,P)\rightarrow \pm \infty$ (for $\epsilon \lessgtr 2$). This is illustrated in Fig.~\ref{regimestatdyn} for various values of the amplitude of the driving, which breaks the $Q-P$ symmetry. In the following, $f\ge0$ is discussed for definiteness; results for $f\le0$ are trivially related by $Q\rightarrow -Q$.

In case of $\epsilon\gg 1$, the relevant regime is the one of small values for $Q^2$ and $P^2$ with a weak anharmonicity. It corresponds locally  to a Duffing oscillator. 
In contrast, for $\epsilon\gtrsim 2$ the full impact of the Bessel function comes into play. Apparently, this latter one is the most interesting domain, since then in the presence of dissipation multiple steady state orbits are accessible \footnote{
Note that we assume a conventional Josephson junction with $E_J>0$ (otherwise, we shift $\phi$ by $\pi$ and procced as before), so that for the Josephson potential, which becomes softer for stronger driving, multiple solutions can only possibly occur for $\epsilon >0$. 
}.  
These are obtained  from the stationary solutions of the equations of motion, i.e.,
\begin{equation}
\frac{dQ}{dt}= - \frac{\tilde{\gamma}}{2} Q - \frac{\partial g}{\partial P},\hspace{0.2cm} \frac{dP}{dt}= - \frac{\tilde{\gamma}}{2} P + \frac{\partial g}{\partial Q}\, 
\label{eomHam}
\end{equation}
similar to \cite{DykmanMaloneySilverstein1998}.

\begin{figure} 
\includegraphics[width=1\linewidth]{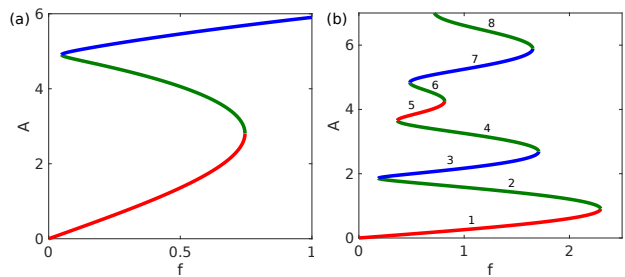} 
\caption{Steady state amplitude $A=\sqrt{Q^2+P^2}/2\kappa$ for (a) $\epsilon=17.3$, $\kappa=0.1$ and (b) $\epsilon=2.05$, $\kappa=1$ both including  dissipation $\tilde{\gamma}=0.05$. For $\epsilon\gg 1$ the bifurcation pattern known from the Duffing oscillator is recovered. For $\epsilon\gtrsim 2$ replicas of the Duffing orbits appear at larger amplitudes. Red segments indicate stable orbits located at the minima of the quasi-energies for $Q \geq 0$ (labels 1 and 5, cf.~Fig.~\ref{regimestatdyn}), while stable orbits corresponding to maxima at $Q<0$ (labeled 3 and 7) are shown in blue. Unstable solutions (green, even labels) refer to maxima (minima) at $Q>0$ ($Q<0$).}
\label{regimestatdyntwo}
\end{figure}

Results for fixed points amplitudes for different values of $\epsilon>2$ are shown in Fig.~\ref{regimestatdyntwo}. As expected, one sees that for $\epsilon\gg 1$ and $f/\epsilon\ll 1$ the bifurcation pattern known from the Duffing oscillator is recovered, while this picture becomes much richer for $\epsilon\gtrsim 2$ and $f/\epsilon\sim O(1)$. There, replicas of the Duffing orbits appear at larger amplitudes. A  stability analysis (see Appendix \ref{sec:stability}) reveals that for values of $  \epsilon \ll 2 $ as well as $ \epsilon\gg 2 $ only a single stable orbit exists around $Q=0$.
Approaching  $ \epsilon = 2 $ from either side, more and more pairs of stable and unstable orbits with larger $|Q|$ successively appear (cf. \cite{CaputoGabitovMaimistov2012, LazaridesTsironis2013, JungUstinov2014, PhysRevB.90.205411,MaizelisRuderDykman2014} for multistable solutions). Specifically, one finds stable orbits at the minima of the quasi-energies in 
Fig.~\ref{regimestatdyn}  for $Q \geq 0$ (labeled 1 and 5 and corresponding to amplitudes  colored red in Fig.~\ref{regimestatdyntwo})  and at the maxima at $Q<0$ (labeled 3 and 7 and colored blue).  All maxima (minima) of $g(Q,P)$  for $Q>0$ ($Q<0$) correspond to unstable orbits (green-colored, labeled 2, 4, 6, 8). 

\section{Bare quasi-energies and eigenfunctions}
\label{barehamilton}
\begin{figure}
\includegraphics[width=1\linewidth]{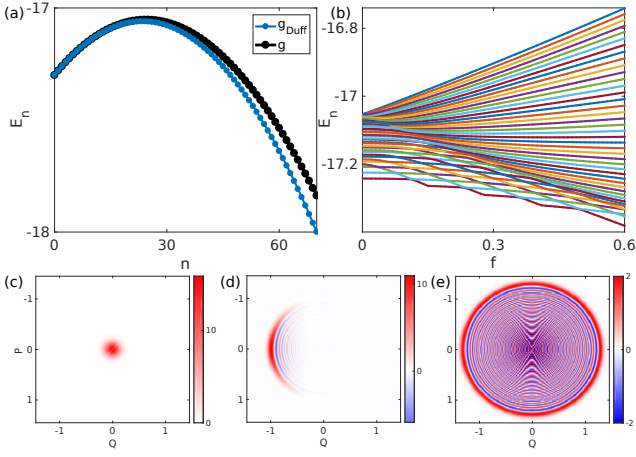} 
\caption{Quasienergy spectrum for the Josephson Hamiltonian $g$ in the rotating frame for $\kappa=0.1$, $\epsilon=17.3$ and increasing $n$; (a) without driving and (b) for changing driving amplitude. The eigenenergies form a parabola with $n$ resembling the classical amplitude dependence. With growing $f$ an ordering of the eigenvalues according to Fock state numbers is no longer possible. In the bottom row Wigner transforms of eigenstates for (c) $n=0$, (d) $n=25$, (e) $n=43$ are plotted (for a minute but finite $f=0.01$). The small driving localizes the state $n=25$ corresponding to the top of the energy parabola at $Q<0$, while the $n=0$ and $n=43$ states remain centered around $Q=P=0$. }
\label{quasidetone}
\end{figure}
\begin{figure}
  \includegraphics[width=1\linewidth]{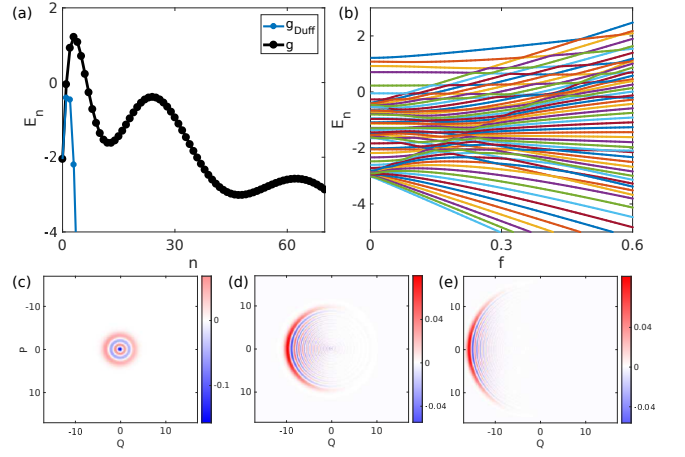}
\caption{Quasienergy spectrum for the Josephson Hamiltonian $g$ for $\kappa=1$, $\epsilon=2.05$ and increasing $n$; (a) without driving and (b) for changing driving amplitude. Instead of the simple parabola-shape of Fig.~\ref{quasidetone}(a) multiple extrema in the quasi-energy spectrum appear for increasing $n$. Wigner transforms of eigenvectors for (c) $n=3$, (d) $n=24$, (e) $n=62$ ($f=0.01$) show for small $n$ a Fock-state-like distribution around $Q=P=0$ for the first maximum. 
Beyond the first energy maximum strongly displaced and squeezed states localized around the maxima at $Q<0$ appear.} 
\label{potentialquasi}
\end{figure}

 
We start by analyzing the quantum properties of the system by considering the bare quasi-energy operator (\ref{clhamrotpot}). 
Without driving $f=0$, the quasi-energy commutes with the number operator $\hat{n}$ and is thus diagonal in the Fock state basis $\hat{n}|n\rangle= n|n\rangle$.  
For small $\kappa$ and large $\epsilon$ a parabola-type spectrum for the quasi-energies $E_n$ results [cf.~Fig.~\ref{quasidetone}(a)], in contrast to the regime of $\kappa\sim 1$ and $\epsilon\gtrsim 2$ [cf.~Fig.~\ref{potentialquasi}(a)], where we see oscillations stemming from the Bessel function similar to Fig.~\ref{regimestatdyn} but with substantial modifications due to the normal-ordering. 

Strictly without driving, $f \equiv 0$, the corresponding eigenstates are simply Fock states. A useful connection between eigenstates and classical solutions can be found, however, by taking a minute but finite driving, $f=0.01$, which breaks the $Q-P$ symmetry. 
In phase space, eigenstates away from maxima or minima of the quasienergy spectrum still resemble the harmonic ones: they are centered around $Q=P=0$ as shown by their Wigner density in Fig.~\ref{quasidetone}(c) and (f). For states around the top of the parabola of eigenenergies [$n\approx25$, Fig.~\ref{quasidetone}(d)], where the energy spacing between neighboring states becomes comparable to the (minute) off-diagonal driving  term, constructive and destructive interference leads to a phase-space localization around the first maximum at $Q<0$ [corresponding to a stable classical orbit, cf. blue line with $A\sim 5$ in Fig.~\ref{regimestatdyntwo}(a)].
The same is seen for states around higher maxima in the strongly nonlinear spectrum [Fig.~\ref{potentialquasi}(d) and (f)]  localized at positions, where higher order maxima in the classical quasi-energy appear. Despite their association with classical solutions the eigenstates can be  strongly squeezed (and displaced) and carry substantial negative parts in the Wigner functions, a signature of their pronounced non-classicality.\\
\begin{figure}[hb]
\includegraphics[width=1\linewidth]{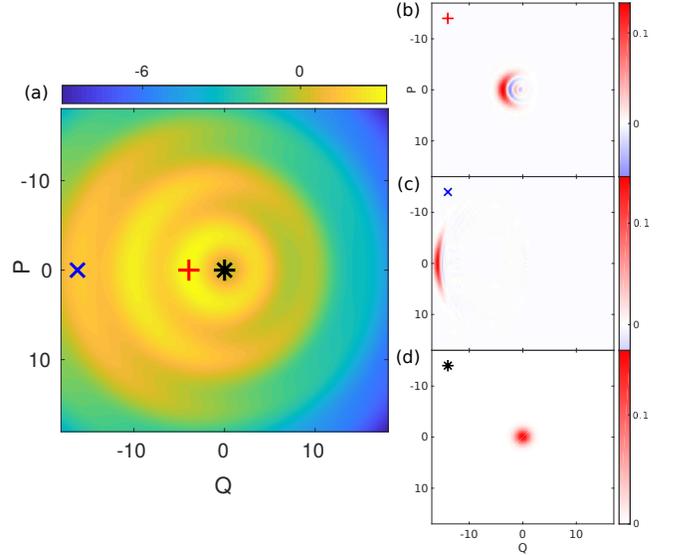}
\caption{Wigner transforms of eigenvectors for (b) $m=0$, (c) $m=14$, (d) $m=48$ (counted in decreasing order of energy) including finite driving $f=0.3$ for $\kappa=1$, $\epsilon=2.05$. Finite driving breaks the $Q-P$ symmetry and thus induces additional displacement and squeezing. 
The eigenstates chosen correspond to classically stable solutions at stationary points in the (classical) quasi-energy landscape (a). For instance, the near-Gaussian state for $m=48$ (d) corresponds to the classical solution in the central well at $Q=P=0$.
} 
\label{wignerfinitef}
\end{figure}

\clearpage
For strong driving, energy eigenstates can no longer be indexed by $n$ (as they are not eigenstates of the number operator), but are ordered by decreasing energy (and for clarity indexed by $m$ in the following). It is illuminating to monitor how spectra evolve with growing $f$ from the ones at $f=0$. For the two $\kappa$-values shown in Fig. \ref{quasidetone}(b) and \ref{potentialquasi}(b), spreading of the eigenenergies with increasing driving is observed. In particular, for $f=0$ there are clusters of energetically-close states around the minima/maxima of Fig. \ref{quasidetone}(a) and \ref{potentialquasi}(a). Even a small driving leads to a strong repulsion among these near-degenerate states. For finite driving, no good quantum number beside the energy remains, so that generically anticrossings appear in the spectra. 

Wigner functions at finite driving are shown in Fig.~\ref{wignerfinitef} for those eigenstates that can be associated with classically stable solutions for that driving strength [Fig.~\ref{wignerfinitef}(a), cf. Fig.~2 and 3(b)].
For instance, the highest-energy eigenstate ($m=0$) corresponds to the stable solution (local minimum) with highest quasi-energy located around $(Q,\,P) = (-4,\,0)$; $m=14$ (c) corresponds to another local minimum at negative $Q\approx -16$ with strong squeezing; and Fig.~6(d) ($m=48$) displays the near-Gaussian eigenstate, which is found around the central well  $Q=P=0$.    
\begin{figure}[t]
\includegraphics[width=1\linewidth]{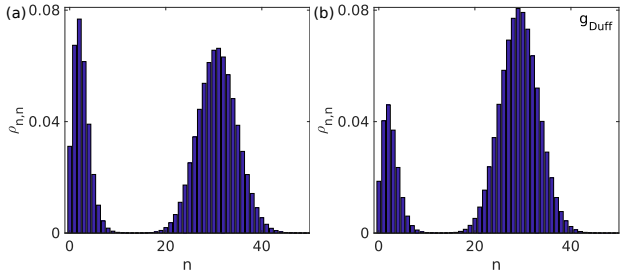}
\caption{Steady state population of the dissipative dynamics described by the dimensionless Lindblad master equation (\ref{master}) for $f=0.56$, $\kappa=0.1$, $\epsilon=17.3$, $\tilde{\gamma}=0.05$. 
For these parameters eigenstate populations of (a) the Josephson Hamiltonian ($\braket{n}=21.89$, max. at $n=31$) are very close to the results of (b) the pure Duffing system ($\braket{n}=24.23$, max. at $n=29$). Both show a bimodal distribution with contributions from the two classically-stable orbits, cf.~Fig.~\ref{regimestatdyntwo}.} 
\label{steadystatesmallkappa}
\end{figure}

\section{Steady state quantum dynamics}
\label{steadystate}
Let us now turn to the  non-equilibrium quantum dynamics.  For this purpose, we employ a master equation formulation at $T=0$ for the dynamics of the reduced density operator of the Josephson system, i.e., 
\begin{equation}
\partial_t {\rho} = -\frac{{i}}{\kappa^2} [g,\rho] + \frac{ \tilde{\gamma} }{2 } (2 {a} \rho {a}^{\dagger} - {n} \rho - \rho {n})\, .
\label{master}
\end{equation}
Steady state solutions of (\ref{master}) are approached in the long time limit and can then be compared either to  corresponding findings for the Duffing case or to the properties of the non-dissipative system. Thereby,  $\kappa$ plays the role of a tuning parameter which allows exploring  quantum-classical transitions and  semiclassical fixed points. Practically, it is convenient to represent the density operator in the Fock state basis of the ground state oscillator, i.e.\ $\rho_{n, m}=\langle n|\rho|m\rangle$. 

\begin{figure}[t]
\includegraphics[width=1.05\linewidth]{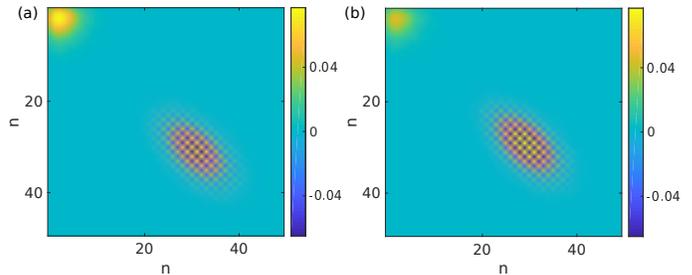}
\caption{Full density  Re$\{\rho_{n,m}\}$ in the steady state for $f=0.56$, $\kappa=0.1$, $\epsilon=17.3$, $\tilde{\gamma}=0.05$ for (a) the Josephson Hamiltonian and for (b) the Duffing system. 
We find two isolated domains in Fock space with the only off-diagonal contributions (coherences) concentrated close the diagonal. The absence of coherences between the two domains indicate a classical mixture of the two classically-stable orbits. The imaginary part of the density shows a similar decrease of coherences with increasing distance from the diagonal.} 
\label{steadystatesmcoh2}
\end{figure}
In the range of small values for $\kappa$ (and large $\epsilon$), one finds steady state eigenstate populations that are very close to the results of the corresponding Duffing system.  Figure~\ref{steadystatesmallkappa} shows results in a regime, where the equivalent classical system relaxes to one of the two stable orbits depending on the initial conditions (cf.~Fig.~\ref{regimestatdyntwo}), while quantum mechanically one always finds a bimodal distribution. This can be understood as mixing of the two classically stable states \cite{GuoZhengLi2010, GuoZhengLiZhengLiYan2011, KatzRetzkerStraubLifshitz2007}. In fact, when one relates in a semiclassical sense the occupation number $n$ to the amplitude $A$ of the classical orbits according to $n\sim A^2$, one finds that the locations of the maxima of the distribution match approximately the stable steady state amplitudes in Fig.~\ref{regimestatdyntwo}(a) at the respective driving, while there is no weight in the distribution at the unstable steady state amplitude. Both peaks are, however, rather broad reflecting quantum fluctuations.
Correspondingly, when inspecting the full density, e.g.,  Re$\{\rho_{n,m}\}$ in Fig.~\ref{steadystatesmcoh2}, one observes two isolated domains in Fock space with off-diagonal contributions (coherences) concentrated close to the diagonal. 
The appearance of a checkerboard sign pattern for the large amplitude solution can be traced back to the fact, that it is located at negative Q-values in phase space. Consequently, there are no sign changes for the small-amplitude situated at positive Q.
There are minor deviations in the location and height of the maxima of the population distributions when comparing the Duffing approximation to results for the full Josephson Hamiltonian. Directly setting the respective quasi-energies for $f=0$ side by side reveals increasing differences for $n$ values beyond the energy maximum, see Fig.~\ref{quasidetone}(a). Overall, however, in this regime the Josephson steady state is well described by making the Duffing approximation.

\begin{figure}[t]
\includegraphics[width=1\linewidth]{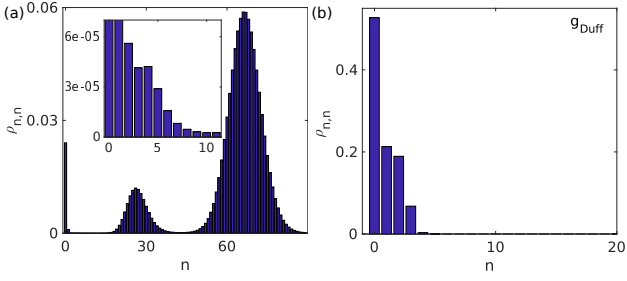}
\caption{Steady state population of the dissipative dynamics for $f=0.8$, $\kappa=1$, $\epsilon=2.05$, $\tilde{\gamma}=0.05$. 
For these parameters, specifically for $\epsilon \sim 2$, full Josephson and Duffing-approximation results differ drastically. The full Josephson population distribution (a) contains more than two maxima, while for the Duffing case (b) only monostability is seen. In that case, the bistability classically expected in this domain is completely washed out by quantum fluctuations. 
The Josephson population distribution exhibits maxima at $n=0$, $n=4$ (see inset), $n=26$, $n=66$.
The higher maxima are related to the maxima in the bare quasi-energy spectrum, Fig.~\ref{potentialquasi}(a).
 Comparing the maxima with the steady state amplitudes in Fig.~\ref{regimestatdyn}, \ref{regimestatdyntwo} reveals that only stable orbits located at $Q\approx 0$ (labeled 1 with $A^2\approx 0.04$) and at  $Q\leq 0$ (labeled 3, 7 with $A^2\approx 4$, and 26, respectively) can be associated with maxima in $\rho_{n,n}$.} 
\label{coherencessmallkap}
\end{figure}

The situation is very different for $\kappa\approx  1$ and smaller $\epsilon$, see Fig.~\ref{coherencessmallkap}.  The full Josephson population distribution in (a) contains in the range $0\leq n\leq 90$ now three pronounced maxima as well as a very weak one around $n=4$ almost hidden behind the strong background of the peak at $n=0$ (see inset). 
Interestingly, maxima in the population distribution for large $n$ may exceed those of smaller $n$. The Duffing case, in contrast,  exhibits only a single peak around $n=0$.
The maxima at $n=4,\,26,\,66$ are clearly related to the maxima in the bare quasi-energy spectrum, see Fig.~\ref{potentialquasi}(a). Comparing the location of the Josephson peaks with the classical steady state amplitudes $A$ in Fig.~\ref{regimestatdyntwo} and the topology of the quasi-energy in Fig.~\ref{regimestatdyn} the following picture appears:  only the central stable orbit located at $Q\approx 0$ (labeled 1 with $A^2\approx 0.04$) and the orbits with $Q\leq 0$ (labeled 3, 7 with $A^2\approx 4,\,26$, respectively) can be associated with maxima in the populations $\rho_{n,n}$, while the one located at $Q>0$ (labeled 5 with $A^2 \approx 17$) seems to be absent. 
Classically, some intuition about the stability of orbits against fluctuations can be gained from quasi-energy plots such as Fig.~\ref{regimestatdyn}, where orbit 5 resides in a very shallow minimum, and bifurcation diagrams, such as 
Fig.~\ref{regimestatdyntwo}, where a driving of $f=0.8$ is just below the critical driving strength at which the orbit disappears. While, generically, the quantum mechanical picture is very complex \cite{MarthalerDykmanswitchin2006,DykmanMarthalerPeano2011}, such intuition about  classical (in)stability nonetheless hints, whether an orbit will carry considerable weight in the steady state distributions or is washed out by fluctuations and thus explains the absence of a peak for orbit $5$ at $n\approx A^2 \approx 17$.

\begin{figure}[!ht]
\hspace{-0.45cm} \includegraphics[width=0.51\linewidth]{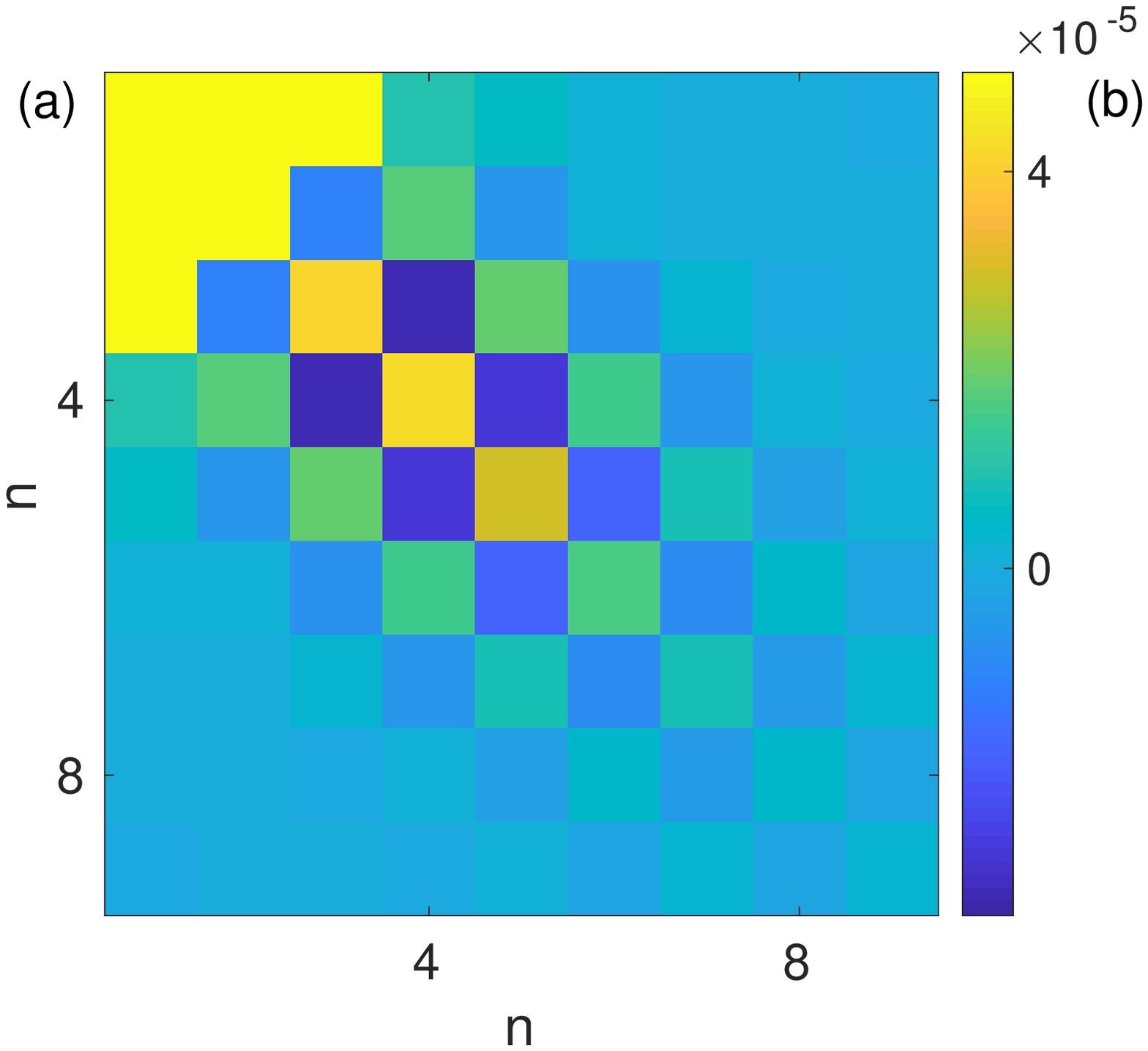}
\hspace{-0.235cm} \includegraphics[width=0.535\linewidth]{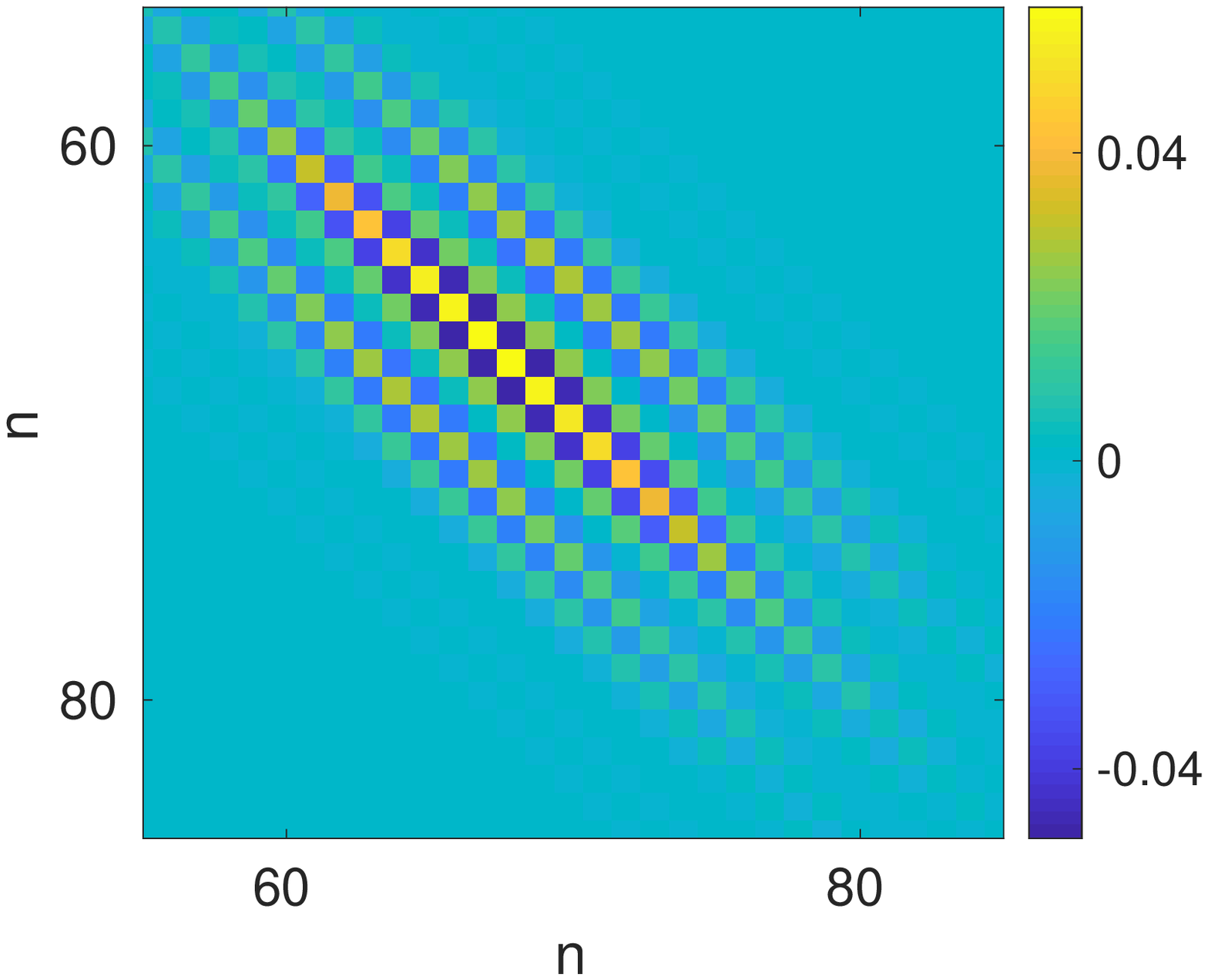} \hspace{-0.9cm} 
\caption{Full density  Re$\{\rho_{n,m}\}$ in the steady state for $f=0.8$, $\kappa=1$, $\epsilon=2.05$, $\tilde{\gamma}=0.05$ for the Josephson Hamiltonian. The Fock space density shows distinct domains according to the three pronounced peaks in Fig.~\ref{coherencessmallkap} with no coherences between them. The right section (b) shows the coherences concentrated close to the diagonal for the largest peak. The left section (a) of small $n$ reveals substantial coherences between the weakly pronounced  peak at $n=4$ and the central $n=0$ contribution.
}
\label{steadystatesmcoh}
\end{figure}

The full Fock space density shown in Fig.~\ref{steadystatesmcoh} displays basically three distinct domains according to the three pronounced peaks with no discernible coherences between the distinct peaks. Only for the weak peak at $n=4$ and the central  $n=0$ peak substantial off-diagonal contributions appear, indicating a type of `hybridization' of two classical steady state orbits. Again the checkerboard pattern appears for solutions located at negative $Q$.\\
~
\vspace*{-0.5cm}\\
To study fluctuations around fixed points for larger amplitudes, a semiclassical analysis is illuminating (cf. \cite{ArmourBlencoweRimberg2013}). 
For this purpose, we introduce  $a \rightarrow \alpha + \delta a$, with $\alpha=A e^{-i \phi}$ being a classical fixed point with amplitude $A=\sqrt{Q^2+P^2}/2 \kappa$ and  phase $\phi={\rm arctan}(\frac{P}{Q})$ while $\delta a $ ($\delta a^\dagger$) denote local annihilation (creation) operators $[\delta a, \delta a^\dagger]=1$. Accordingly, one considers a displaced reduced density operator $\sigma=D^\dagger(\alpha) \rho D(\alpha)$ with the displacement operator $D(\alpha)=\exp(\alpha a^\dagger-\alpha* a)$ and neglects higher than second order terms in the quasi-energy operator (\ref{clhamrotpot}), i.e.\ $g\to g_2$, with
\begin{eqnarray}
g_2& =&  \kappa^2\, [2-\epsilon+\epsilon J_0(2\kappa |\alpha|)] \delta a^\dagger  \delta a \notag \\
&&-\epsilon \kappa^2 \frac{J_2(2\kappa|\alpha|)}{2  |\alpha|^2} [(\alpha^*)^2 (\delta a)^2+\alpha^2 (\delta a^\dagger)^2]
\end{eqnarray}
with the Bessel functions $J_0$ and $J_2$. In particular, one has $\langle n\rangle = |\alpha|^2=A^2$ and $\langle \delta a\rangle =0$ and $\sigma$ follows from (\ref{master}) with $g$ replaced by $g_2$.  Accordingly, the Heisenberg equations for the displaced operators read

\begin{figure}
\includegraphics[width=1\linewidth]{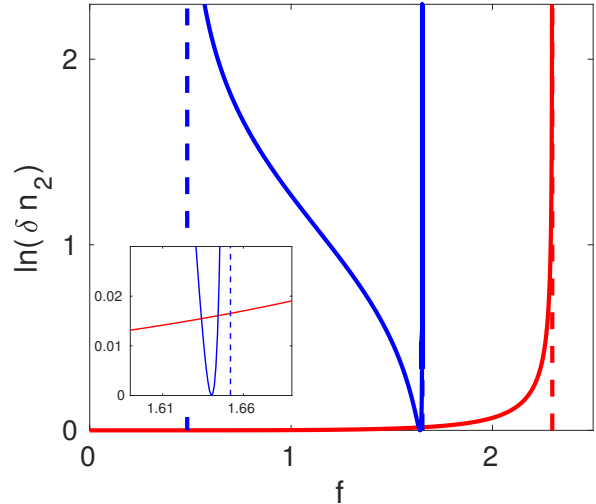}
\caption{Number fluctuations for $\kappa=1$, $\epsilon=2.05$, $\tilde{\gamma}=0.05$ around the semiclassical solution with $n\approx 0$ (red) and $n\approx 26$ (blue) for different driving strengths. The red (blue) solid line can be associated with the red branch of smallest amplitude (the branch with $A\approx 5$) in Fig.~\ref{regimestatdyntwo}(b).
The number fluctuations for $n \approx 0$ take the harmonic value, $\delta n_2=1$, for small driving. With a higher driving the number fluctuations grow and eventually diverge at the bifurcation point. Number fluctuations for $n\approx 26$ grow for a small driving and at a stronger driving amplitude, because on both sides bifurcation points are present. However, between these regions a narrow domain is found, where the system reduces again to a coherent state with $\ln(\delta n_2)=0$.} 
\label{fluct}
\end{figure}

\begin{widetext}
\begin{equation}
\begin{pmatrix}
\delta \dot{a}\\
\delta \dot{a}^\dagger
\end{pmatrix}\\=
\begin{pmatrix}
-i[2 -\epsilon  +\epsilon J_0(2\kappa |\alpha|)]- \frac{\tilde{\gamma}}{2}  & {i \epsilon \alpha^2} J_2(2\kappa|\alpha|)/|\alpha|^2\\
-
{i \epsilon (\alpha^*)^2} J_2(2\kappa|\alpha|)/|\alpha|^2  &i[2 -\epsilon  +\epsilon J_0(2\kappa |\alpha|)]- \frac{\tilde{\gamma}}{2} 
\end{pmatrix}
\begin{pmatrix}
\delta a \\
\delta a^\dagger 
\end{pmatrix} + \sqrt{\tilde{\gamma}}
\begin{pmatrix}
\delta a_{\rm in} \\
\delta a^\dagger_{\rm in} 
\end{pmatrix}
\label{deltamatmat}
\end{equation}
\end{widetext} 
where the noise operators $a_{\rm in},a_{\rm in}^\dagger$ obey: $\langle a_{\rm in}\rangle=\langle a_{\rm in}^\dagger\rangle=0 $, $\langle a_{\rm in}(t)a_{\rm in}(t') \rangle$=$\langle a_{\rm in}^\dagger(t)a_{\rm in}^\dagger(t') \rangle$=$\langle a_{\rm in}^\dagger(t)a_{\rm in}(t') \rangle=0$ and $\langle a_{\rm in}(t)a_{\rm in}^\dagger(t') \rangle=\delta(t-t')$. Note that here $\alpha$ is a function of the parameters $f$ and $\epsilon$. 
This way, one finds for the number fluctuations the expression
\begin{eqnarray}
\label{numberfluc}
\delta n_2 &\equiv &\langle \delta a \delta a^\dagger +\delta a^\dagger \delta a\rangle \\
&=&\frac{[2 -\epsilon  +\epsilon J_0(2\kappa |\alpha|)]^2+\tilde{\gamma}^2/4}{[2 -\epsilon  +\epsilon J_0(2\kappa |\alpha|)]^2+\tilde{\gamma}^2/4-\epsilon^2 J_2(2\kappa|\alpha|)^2}
\, \nonumber.
\label{deltamat}
\end{eqnarray}
The eigenvalues of the matrix  in (\ref{deltamatmat}) determine the stability of the fixed point solutions and are obtained as
\begin{equation}
\hspace{0.03cm} \lambda_\pm = -\frac{\tilde{\gamma}}{2}\pm \left\{\epsilon^2 J_2(2\kappa |\alpha|)^2-[2-\epsilon+\epsilon J_0(2\kappa|\alpha|)]^2\right\}^{1/2}\hspace{-0.3cm} . \hspace{-0.1cm} 
\end{equation}
In the harmonic limit, this reduces to $\lambda_\pm=-\gamma/2\pm 2 i $ and fixed points are always stable. Outside, a change of stability occurs when  $\lambda_+=0$ by tuning $f$ (and thus $|\alpha|$) and/or $\epsilon$ which implies that number fluctuations (or energy fluctuations) according to (\ref{numberfluc}) diverge. Figure \ref{fluct} illustrates this behavior for fixed points corresponding to maxima at $n=0$ and $n=26$ in Fig.~\ref{coherencessmallkap}; in Fig.~\ref{regimestatdyntwo}(b) these may roughly be associated with the classical orbits of least amplitude  (label 1 in Fig.~\ref{regimestatdyntwo}, red line in Fig.~\ref{fluct}) and with $A\approx 5$ (label 5 in  Fig.~\ref{regimestatdyntwo}, blue line in Fig.~\ref{fluct}). For the fixed point emerging from the harmonic one with increasing driving, number fluctuations start from the result for a purely coherent state $\delta n_2=1$ and remain close to this value for a broad range of driving strengths. Close to the instability point $\delta n_2$ sharply grows indicating strong quadrature squeezing. The classical fixed point with higher amplitude exhibits two instability points and possesses only a narrow range of driving, where $\delta n_2$ is close to the coherent state value; outside strong quadrature squeezing dominates. Qualitatively, this originates from the fact that around a bifurcation at $f_c$ with amplitude $A_c$  number fluctuations scale as $\delta n_2\sim A_c/|f-f_c|^\beta$ with $|A-A_c| \propto |f-f_c|^\beta$ with $\beta<1$. In the present case, one has $\beta\approx 1/2$ (cf. Fig.~\ref{regimestatdyntwo}) which implies that the range of $f$-values with $\delta n_2>1$ becomes very broad for larger amplitudes $A_c$. However, between these regions narrow domains are found, where the system reduces again to a coherent state with $\delta n_2=1$. This is a consequence of $J_2(2\kappa|\alpha|)=0$ in \eqref{numberfluc}.

\section{Conclusion}
\label{conclusion}
The quantum dynamics of a Josephson junction driven by an external time-periodic force coupling to its phase is investigated in the regime, where driving is so strong that the conventionally employed reduction to a Duffing oscillator fails. Since there is no static force (dc-current) applied, known resonances in form of Shapiro steps do not exist \cite{Shapiro1963}. Starting from the Hamiltonian in the laboratory frame, we investigate the system in a rotating frame and reveal under which conditions simplifications such as, for example, the Duffing approximation are valid. Ground state fluctuations of the phase induce a renormalization of the bare Josephson coupling, a phenomenon that has been recently detected in a somewhat different setting \cite{Joyez2013,1810.06217}. The junction's response is thus resonant only if the driving frequency matches the renormalized plasma frequency of the junction. Further, the quantum dynamics in the rotating frame and in presence of dissipation is studied within a master equation approach. Results are discussed in comparison to classical steady state orbits and quantum results for the Duffing system. The main difference to the latter is the occurrence of a complex pattern of multiple fixed points in the classical limit leading quantum mechanically to multiple maxima in the population distribution. Coherences between classically well separated orbits are absent while they are found to be strong for closely adjacent orbits. While classically the system locks to specific fixed points with narrow domains, where changes of stabilities due to bifurcations occur, quantum mechanically, substantially enhanced number fluctuations (or energy fluctuations) corresponding to strong quadrature squeezing are seen over broad ranges of driving strengths. In very narrow domains, the system collapses to coherent states again with strongly suppressed fluctuations though. 
In the laboratory frame large energy fluctuations induced by time-dependent currents manifest themselves as large fluctuations of ac-voltages and thus may be accessible experimentally. Our study sheds light on Josephson physics in a regime that has not been explored yet in depths, in some contrast to the fact that Josephson junctions serve as paradigmatic nonlinear elements for mesoscopic superconducting circuits. It may allow to elucidate transitions between the well-known Duffing system and higher order nonlinear oscillators and may thus contribute to the ongoing experimental activities to explore nonlinear quantum dynamics far from equilibrium.

\section*{acknowledgments}
We are grateful to L.Guo for discussions and for financial support by the German Science Foundation through SFB/TRR 21 and the Center for Integrated Quantum Science and Technology (IQST).


\appendix

\section{Transformation of the Josephson Hamiltonian into the rotating frame}
We start with the Josephson Hamiltonian
\begin{equation}
\begin{split}
H_{\text{JJ}}=& \frac{p_{\varphi}^2}{2M} - M\omega^2 \cos(\varphi) - F \cos(\omega_{\text{d}} t) \varphi\\
 =& \frac{q^2}{2 C} - E_J \cos(\varphi) - F \cos(\omega_d t) \varphi . 
\end{split}
\end{equation}
We transform the above Hamiltonian to a frame rotating with the frequency of the external drive, i.e., ${H}_{}^{\text{RF}}={U}{H}{U}^{\dagger} + i \hbar {U}^{\dagger}(\partial_t  {U})$ with the unitary operator ${U}=e^{i \omega_{\text{d}} t {a}^{\dagger}{a}} $. Here $a, a^\dagger$ denote canonical annihilation and creation operators, respectively, according to $\hat{Q}= \kappa (a+a^\dagger)$ with $[a,a^\dagger]=1$ and ground state width of the Josephson potential $\kappa^2=\sqrt{E_C/2 E_J}$  with charging energy $E_C=2 e^2/C$.

By applying the unitary transformation operators to the nonlinear parts of the Josephson Hamiltonian
\begin{equation}
\begin{split}
&e^{i \omega_\text{d} t {a}^{\dagger} {a}} e^{\pm i \sqrt{\frac{\hbar}{2M \omega}}({a}+{a}^{\dagger})} e^{-i \omega_\text{d} t {a}^{\dagger} {a}}\\
&=  e^{ \pm i \sqrt{\frac{\hbar}{2M \omega}}{a}^{\dagger} e^{i \omega_\text{d} t}} e^{\pm i \sqrt{+ \frac{\hbar}{2M \omega}}{a} e^{-i \omega_\text{d} t} } e^{ \frac{\hbar}{2M\omega} \frac{1}{2}[{a}^{\dagger}e^{i \omega_\text{d} t}, {a}e^{-i \omega_\text{d} t}]}\\
&= \sum_{k,u}  \frac{\left(\pm i \sqrt{\frac{\hbar}{2M \omega}}\right)^k}{k!} e^{i \omega_\text{d} t k} ({a}^{\dagger})^{k} \frac{\left(\pm i \sqrt{\frac{\hbar}{2M \omega}}\right)^u}{u!} e^{-i \omega_\text{d} t u} {a}^u e^{-\frac{\hbar}{4M\omega}}\\
&\overset{\text{RWA}}{\underset{\text{k=u}}{=}} :\sum_{k} \frac{\left(- \frac{\hbar}{2M \omega}\right)^k}{k!^2} ({a}^{\dagger} {a} )^k  e^{\frac{-\hbar}{4M\omega}}:\\
&= e^{-\frac{\kappa^2}{2} } :J_0\left(2 \kappa \sqrt{\hat{n}}\right):
\end{split}
\label{waerhfgsf}
\end{equation}
we used the Baker-Campbell-Hausdorff-formula $e^{X+Y}=e^X e^Y e^{-\frac{[X,Y]}{2}}$ and the Bessel function $J_0(z)=\sum_{k=0}^{\infty} (-1)^k (\frac{z}{2})^{2k} \frac{1}{(k!)^2}$.
After the transformation we find the Hamiltonian in the rotating frame 
\begin{equation}
\begin{split}
{H}^\text{RF}_\text{JJ}=& \hbar \bigl(\frac{\omega}{2} - \omega_{\text{d}} \bigr) \hat{n} + \frac{\hbar \omega}{4} + M\omega^2 - M \omega^2 e^{-\frac{\hbar}{4 M \omega}} :J_0 \bigl(2 \sqrt{\frac{\hbar \hat{n}}{2 M \omega}} \bigr):\\
& - \frac{F}{2} \sqrt{\frac{\hbar}{2M \omega}} ({a} + {a}^\dagger)\\
=& \hbar \delta\!\omega \hat{n} - \frac{F}{2} \kappa (a +a^{\dagger}) -E_J^* \left[:J_0\left(2 \kappa \sqrt{\hat{n}}\right):+ \kappa^2 \hat{n}\right].
\end{split}
\end{equation}
To investigate higher orders it is more convenient to write the Bessel function 
\begin{equation}
:J_0\left(2 \kappa \sqrt{\hat{n}}\right):\,=\,:\sum_{k=0}^\infty \frac{\left(- \kappa^2 \right)^k}{k!^2} (\hat{n})^k  :\, =\sum_{k=0}^\infty \frac{\left(- \kappa^2 \right)^k}{k!^2} \prod_{t=0}^{k-1} (\hat{n}-t)
\end{equation}
including a product of $\hat{n}$, where a normal ordering is not necessary any more.\\

\section{Classical stability analysis\label{sec:stability}}
To perform a stability analysis of the steady state solutions of
\begin{equation}
F_1 \coloneqq \frac{dQ}{dt}= - \frac{\tilde{\gamma}}{2} Q - \frac{\partial g}{\partial P}
\end{equation}
and
\begin{equation}
F_2 \coloneqq \frac{dP}{dt}= - \frac{\tilde{\gamma}}{2} P + \frac{\partial g}{\partial Q}\, .
\end{equation}
we compute the extrema for the Jacobi Matrix
\begin{equation}
J =  \begin{pmatrix}
\frac{\partial F_1}{\partial X} & \frac{\partial F_1}{\partial P} \\
\frac{\partial F_2}{\partial X} & \frac{\partial F_2}{\partial P}
\end{pmatrix}
\end{equation}
using
$\frac{\partial}{\partial X} (\frac{J_1(\Delta)}{\Delta})= \frac{\partial}{\partial \Delta} (\frac{J_1(\Delta)}{\Delta}) \frac{\partial \Delta}{\partial X}=- \frac{J_2(\Delta)}{\Delta} \frac{\partial \Delta}{\partial X}$.
Stable solutions are related to a negative real part of the eigenvalues, unstable ones to a positive real part of the eigenvalues. 

\bibliography{literature}
\bibliographystyle{apsrev4-1}

\end{document}